\documentstyle[aps,prl,amssymb,epsfig]{revtex}

\def\dpar#1#2{{\partial #1 \over \partial #2}}

\def\lam {\lambda}

\begin{document}
\draft
\twocolumn[\hsize\textwidth\columnwidth\hsize\csname %
@twocolumnfalse\endcsname
\begin{flushright}
Preprint CAMTP/97-5\\
October 1997
\end{flushright}
\title{Energy level statistics in the transition regime between integrability
and chaos for systems with broken antiunitary symmetry}
\author{Marko Robnik$\,^{1}$, Jure Dobnikar$\,^{2}$ and 
Toma\v z Prosen$\,^{3}$}
\address{$\,^{1}$ Center for Applied Mathematics and Theoretical Physics,
University of Maribor, SI-2000 Maribor\\
$\,^{2}$ Jozef Stefan Institute, SI-1001, Ljubljana\\
$\,^{3}$ Faculty of Mathematics and Physics, University of Ljubljana, 
SI-1000, Ljubljana\\
e-mails: robnik@uni-mb.si, jure.dobnikar@ijs.si, prosen@fiz.uni-lj.si}

\maketitle

\begin{abstract}
\widetext
\smallskip
Energy spectra of a particle with mass $m$ and charge $e$ in the cubic
Aharonov-Bohm billiard containing around $10^4$ consecutive levels
starting from the ground state have been analysed. The cubic
Aharonov-Bohm billiard is a plane billiard defined by the cubic
conformal mapping of the unit disc pervaded by a point magnetic flux
through the origin perpendicular to the plane of the billiard. The
magnetic flux does not influence the classical dynamics, but breaks
the antiunitary symmetry in the system, which affects the statistics
of energy levels.  By varying the shape parameter $\lam$ the classical
dynamics goes from integrable ($\lam =0$) to fully chaotic ($\lam =
0.2$; Africa billiard). The level spacing distribution $P(S)$ and the
number variance $\Sigma^{2}(L)$ have been studied for 13 different
shape parameters on the interval ($0\le\lam\le0.2$). GUE statistics
has proven correct for completely chaotic case, while in the mixed
regime the fractional power law level repulsion has been observed. The
exponent of the level repulsion has been analysed and is found to
change smoothly from 0 to 2 as the dynamics goes from integrable to
ergodic. Further on, the semiclassical Berry-Robnik theory has been
examined. We argue that the semiclassical regime has not been reached
and give an estimate for the number of energy levels required for the
Berry-Robnik statistics to apply.
\end{abstract}
\pacs{PACS numbers: 05.45.+B, 05.40.+J, 03.65GE, 03.65.-W}
]
\narrowtext

The objective of this paper is to study the energy level statistics of
a system without antiunitary symmetry. Introducing magnetic field to a
plane billiard violates the time reversal symmetry. However, if there
are space symmetries present, there can be a combined space-time
symmetry in the system, which is antiunitary
\cite{antiunit}. Therefore, a billiard with no space symmetries is
required.  The natural choice is the cubic billiard defined by the
conformal mapping of the unit disc ($\vert z\vert \le 1$)
\begin{equation}
 w=z+\lam z^2+\lam e^{i{\textstyle{\pi\over 3}}} z^3 
\label{conf}
\end{equation}
onto the physical plane $w$, with the shape parameter $\lam$. A point
magnetic flux through the origin perpendicular to the plane of the
billiard is introduced. Such billiard systems are called Aharonov-Bohm
billiards.  The classical dynamics is unaffected by the point
flux{\footnote {Only the trajectories which pass through the origin
exactly are affected, but they have measure zero among the
trajectories.}} and the system has the scaling property, which says
that the classical dynamics is the same at all energies. The particle
is therefore considered to have a unit speed. This family of billiards
was introduced and first studied by Berry and Robnik \cite{cubic}.

Energy level statistics in the limiting, fully chaotic case of
Africa billiard ($\lambda=0.2$), has been found to be consistent
with the Gaussian Unitary Ensembles (GUE) of random matrices
\cite{R92,BGS}. In the present work we have investigated the energy 
level statistics in the regime of mixed classical dynamics and have
found {\sl the fractional power law level repulsion}{\footnote {The
phenomenon of fractional power law level repulsion has been typically
found in mixed systems with antiunitary symmetry and not too high
energies (see \cite{PR94,PR93} and the references therein).}}, while
we argue that the semiclassical theory of Berry and Robnik \cite{BR}
applies for much larger sequential quantum numbers (see also
\cite{RP97}), in our case say $10^8$. We also reconfirm
\cite{cubic,R92} the quadratic level repulsion and the validity of GUE
(level spacing distribution and number variance) statistics in the
ergodic case with high statistical accuracy.
\\\\
The Schr\"odinger equation for the system \cite{cubic} is
\begin{eqnarray}
\nabla^2\psi(r,\phi)-{2i\alpha \over r^2} \dpar{\psi} {\phi}-
{\alpha^2 \over r^2} \psi(r,\phi)&+&
k^2\left\vert {dw\over dz} \right\vert ^2\psi(r,\phi)=0\;,\nonumber
\\
\psi(1,\phi)&=&0\;.
\label{schro}
\end{eqnarray}
Here $k^2={2mE\over \hbar}$ where $E$ is the energy and the parameter
$\alpha={e\Phi\over 2\pi\hbar}$, where $\Phi$ is the magnitude of the
magnetic flux and $2\pi\hbar$ is the Planck's constant. By
$(r,\phi)$ we denote the polar coordinates in the $z$ plane. The
relevant region of the values of $\alpha$ is $\lbrack
0,{\textstyle{1\over 2}}\rbrack$. It can easily be shown
\cite{antiunit,cubic,R92} that the antiunitary symmetry is still 
present for integer and half integer values of $\alpha$, so the most
appropriate value is $\alpha={1\over 4}$.  We choose exclusively this
value in our present calculations.  In the integrable case ($\lam =0$)
we expect to find Poissonian energy level distribution
\begin{equation}
P(S)=e^{-S}\;,
\end{equation}
while in the completely chaotic case $P(S)$ is predicted by the GUE of
the random matrix theory. The Wigner approximation {\footnote {We
tested the quality of the Wigner approximation by comparing it to the
exact infinitely dimensional solution (see \cite{Haake}) and found
that Wigner approximation is almost perfectly correct, the tiny
deviations from the exact solution are two orders of magnitude smaller
than the deviations of our numerical results.}} gives for GUE
\begin{equation}
P(S)={\textstyle {32\over \pi^2}} S^2 e^{-{\textstyle{4\over \pi}}
S^2}\;.
\label{GUE}
\end{equation}
For small $S$ we have $P(S)\propto S^2$, i.e.  quadratic level
repulsion.

Let us first describe a semiclassical argument (see \cite{BohTomUl})
to estimate the number of converged energy levels in our
calculation. By ``converged'' we mean the accuracy of at least one
percent of the mean level spacing. We expressed the matrix elements of
the Hamiltonian $H$ in a basis defined by eigenvectors of the
integrable Hamiltonian $H^0$ (circular billiard, $\lam = 0$, see
\cite{cubic,diploma}). We truncated the basis at $N^0$th vector and
diagonalised the finite ($N^0\times N^0$) matrix. The eigenvectors
$\vert n\rangle$ of $H^0$ are defined as $H^0\vert n\rangle=e_n\vert
n\rangle$ and the eigenvalues of $H$ as $H\vert
\alpha\rangle=E_{\alpha}\vert \alpha\rangle$. The question is, how
many accurate energy levels $E_{\alpha}$ associated with $H$ are there
in such calculation? In the semiclassical limit the Wigner transform
of an eigenstate with energy $E$ is localized on the classical energy
surface (see \cite{R97,GMW} and the references therein).
\begin{figure}
\epsfig{file=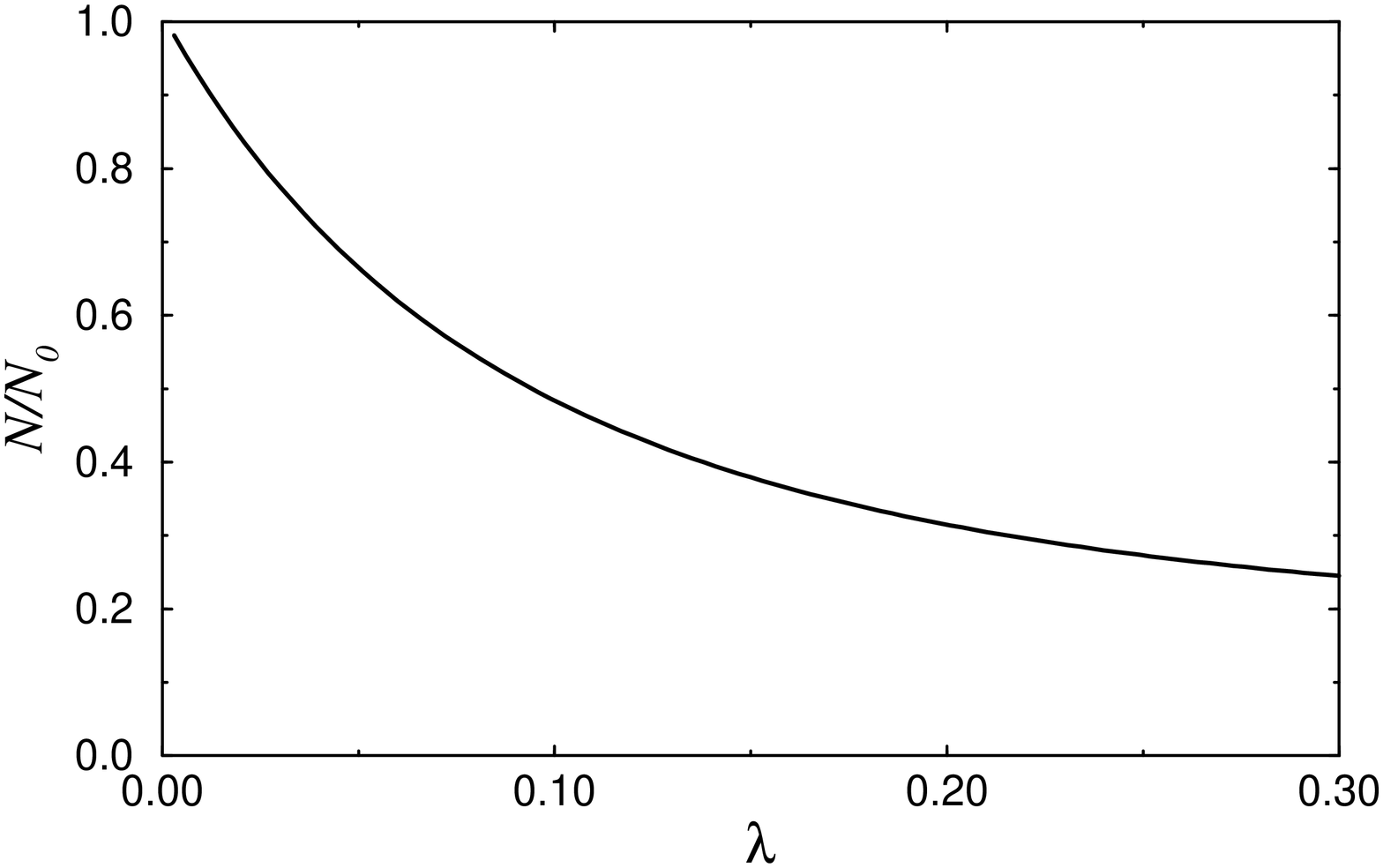,height=8cm,width=4.65cm,angle=-90}
\caption{Number of converged levels as a function of the shape 
parameter $\lambda$.}
\label{fcel}
\end{figure}
This yields that an energy level $E_{\alpha}$ of $H$ will converge if
all eigenvectors $\vert n\rangle$ of $H^0$ with the energy surfaces
$H^0(\vec{q},\vec{p})=E^0_n$ intersecting the energy surface of $H$,
$H(\vec{q},\vec{p})=E_\alpha$, are present in the truncated
basis. {\footnote {It can be shown that the terms in the Hamiltonian
(\ref{schro}) containing $\alpha$ are of the second order in $\hbar$,
so the magnetic flux does not influence the density of states in the
semiclassical limit (of course, it mixes the levels and affects the
statistics).}} Geometrical consideration of foliation of energy
surfaces (see \cite{R97} and the references therein) gives
\begin{equation}
{N\over N^0}=r(\lambda)={A(\lam)\over \pi\; \max_{\vert z\vert \le 1}
\left\vert {dw\over dz} \right\vert ^2}\; ,
\label{NoGL}
\end{equation}
where $A(\lam)$ is the area enclosed by the billiard boundaries and is
equal to $A(\lam)=\pi(1+5\lam^2)$ for the cubic billiard. The fraction
of the converged levels as a function of $\lam$ is plotted in Fig.
\ref{fcel}. Approximately one third of the levels are correct
at $\lam=0.2$, whereas at $\lam=0.05$ there are about two thirds of
good levels. As mentioned above, by a good level we mean the accuracy
of at least one percent of the mean level spacing. Of course, this a
priori criterion is helpful, but the final judgement of the precision
was based on the actual numerical convergence of the levels.

If the basis functions are arranged in a smart way, the $N^0\times N^0$
matrix which has to be diagonalised has band structure. In fact we do not
diagonalise the Hamiltonian, but its inverse $H^{-1}$ \cite{R84} which
can have the desired band structure. The rearrangement of the basis is
described in \cite{PR93}. The CPU time demand of the diagonalisation
drops from ${\cal O}(N^3)$ to ${\cal O}(N^{2.5})$ due to the
rearrangement. The size of the matrix was $N^0=18000$ for $\lam=0.2$
and $N^0=15500$ for thirteen other shape parameters from the Table
\ref{tab0}. By diagonalising the band matrices the fourteen spectra 
have been obtained containing $r(\lambda)N^0$ accurate levels. 

\begin{figure}
\epsfig{file=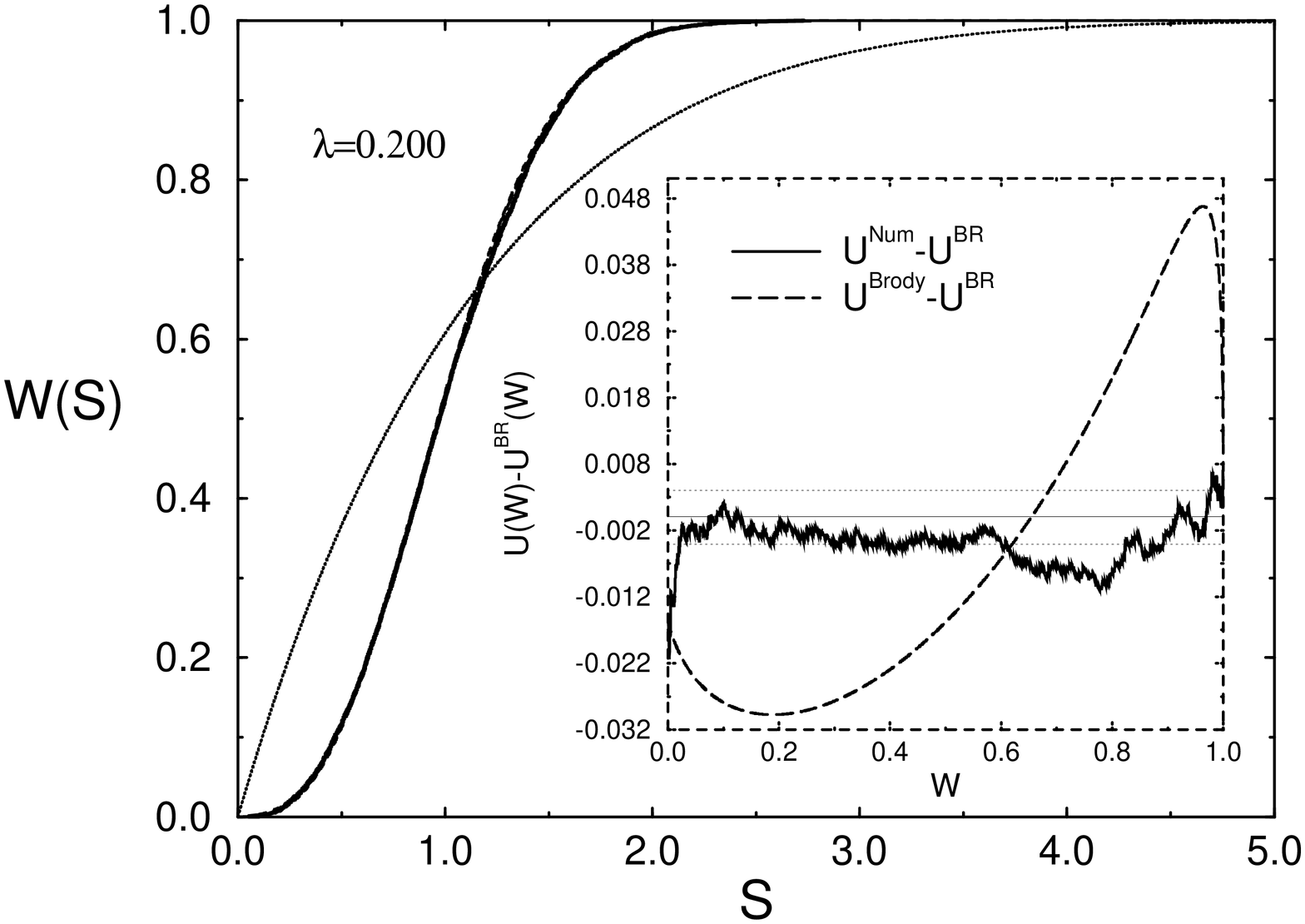,height=8cm,width=3.40cm,angle=-90,
                        bbllx=80,bblly=80,bburx=520,bbury=770}
\vspace{0.5 cm}
\caption{The cumulative level spacing distribution $W(S)$. Numerical 
data: full line, Poisson: dotted line, GUE: dashed line and
Berry-Robnik: dash-dotted line. For the definition of the $U$
functions plotted in the insets, see text. }
\label{WS}
\end{figure}

\begin{figure}[H]
\epsfig{file=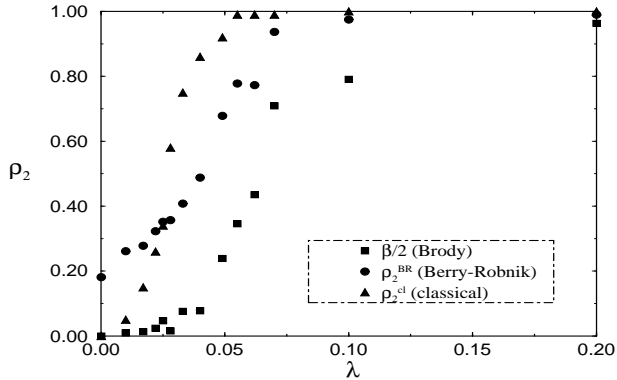,height=8cm,width=5.0cm,angle=-90}
\vspace{0.5 cm}
\caption{Calculated values of $\rho_2^{cl}$, $\rho_2^{BR}$ and 
${\textstyle{1\over 2}}\beta_{Brody}$ as functions of $\lambda$.}
\label{final}
\end{figure}

Using the Weyl formula the levels have been unfolded to unit mean
level spacing and nearest neighbour level spacing distribution $P(S)$
and number variance $\Sigma^2$ have been calculated. Instead of $P(S)$
the cumulative level spacing distribution $W(S)=\int_0^{S}P(s)ds$ is
shown in Fig. \ref{WS}. There are the so-called $U$-functions
\cite{PR94,PR93} shown in the insets. By fitting data {\sl locally} 
($S\le 0.5$) to the Brody distribution,
$P^{\beta}(S)=aS^{\beta}\exp(-bS^{\beta +1})$
\cite{Brody}, the behaviour of $P(S)$ for small $S$ in the transition
region $0\le\lam\le0.2$ has been analysed. However, we know that the
Brody distribution can not {\sl globally} (for all $S$) capture the
physical $P(S)$ (at least for $\beta >1$). Nevertheless, we found a
good agreement with the Brody distribution for $S<0.5$. 
\begin{figure}
\epsfig{file=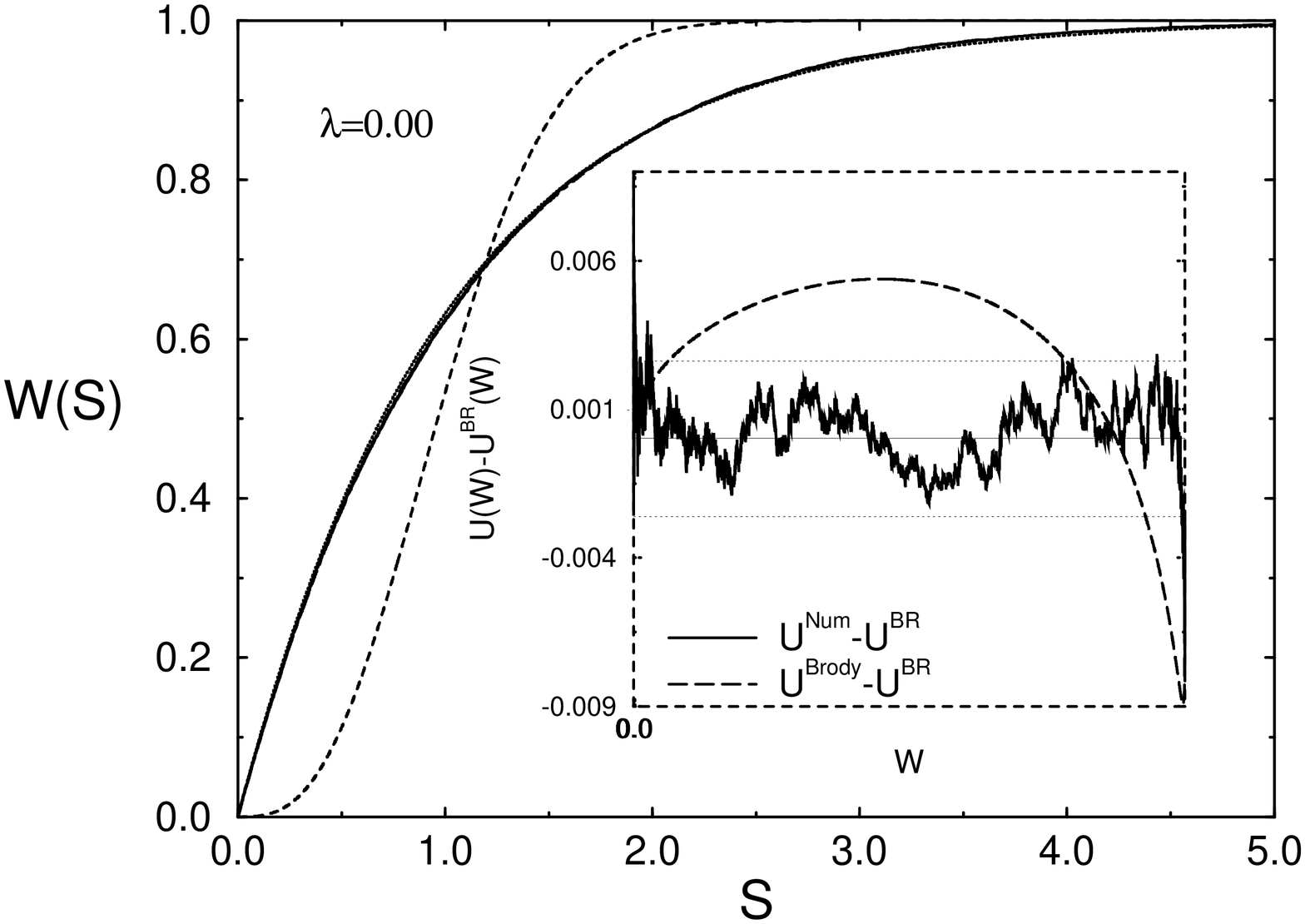,height=8cm,width=3.40cm,angle=-90,
                        bbllx=80,bblly=80,bburx=520,bbury=770}
\end{figure}
\begin{figure}
\epsfig{file=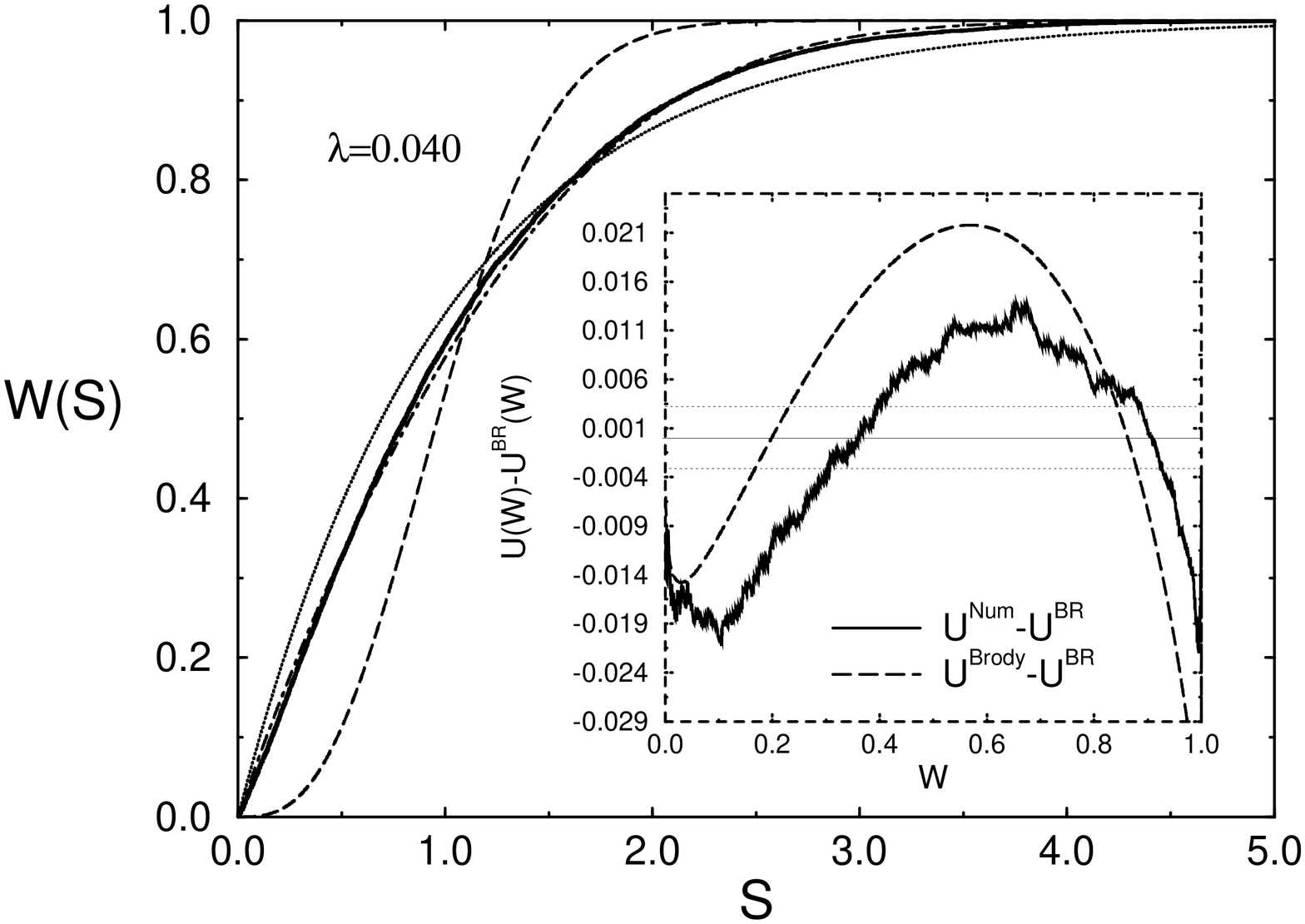,height=8cm,width=3.40cm,angle=-90,
                        bbllx=80,bblly=80,bburx=520,bbury=770}
\end{figure}
\begin{figure}
\epsfig{file=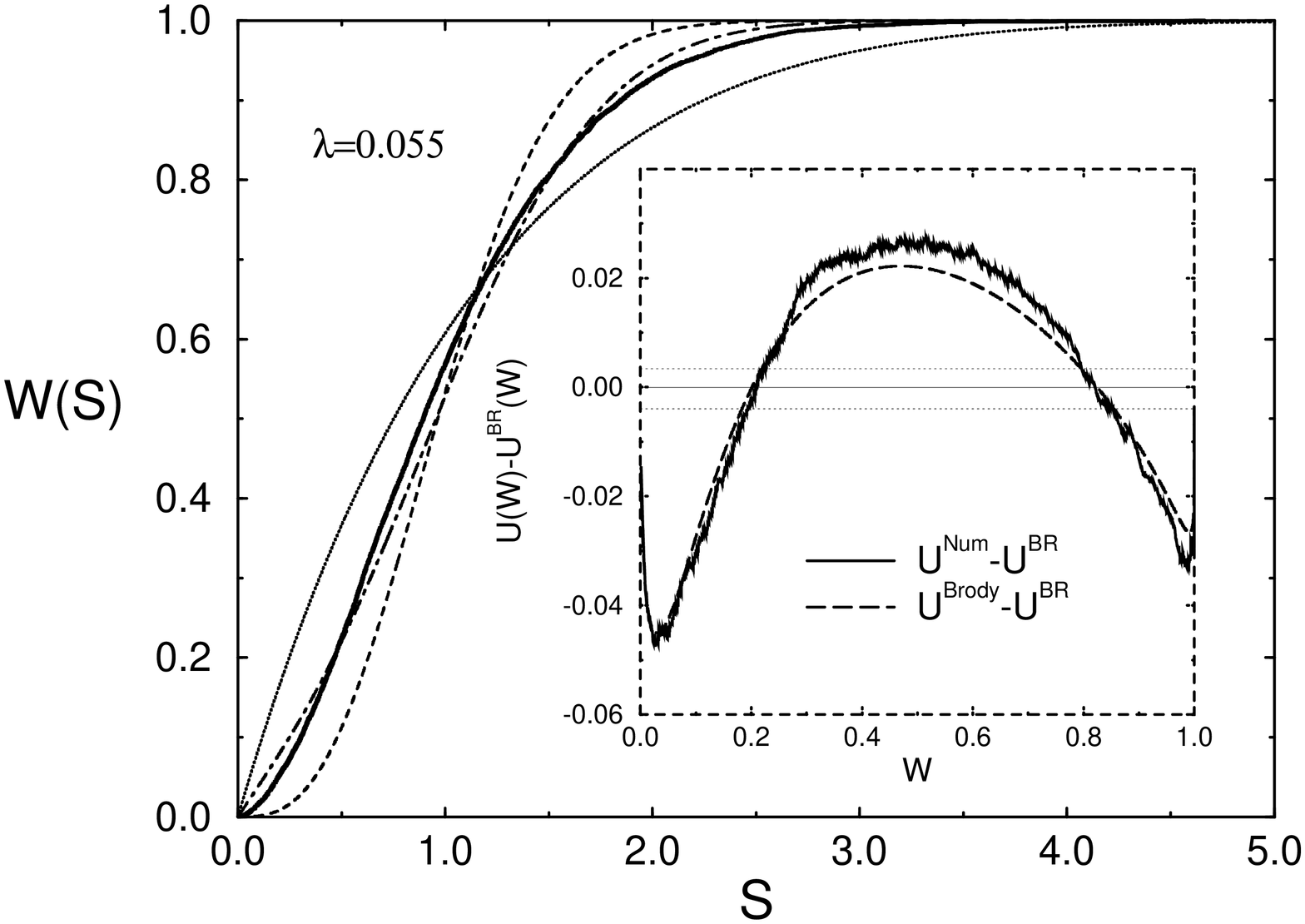,height=8cm,width=3.40cm,angle=-90,
                        bbllx=80,bblly=80,bburx=520,bbury=770}
\end{figure}
\begin{figure}
\epsfig{file=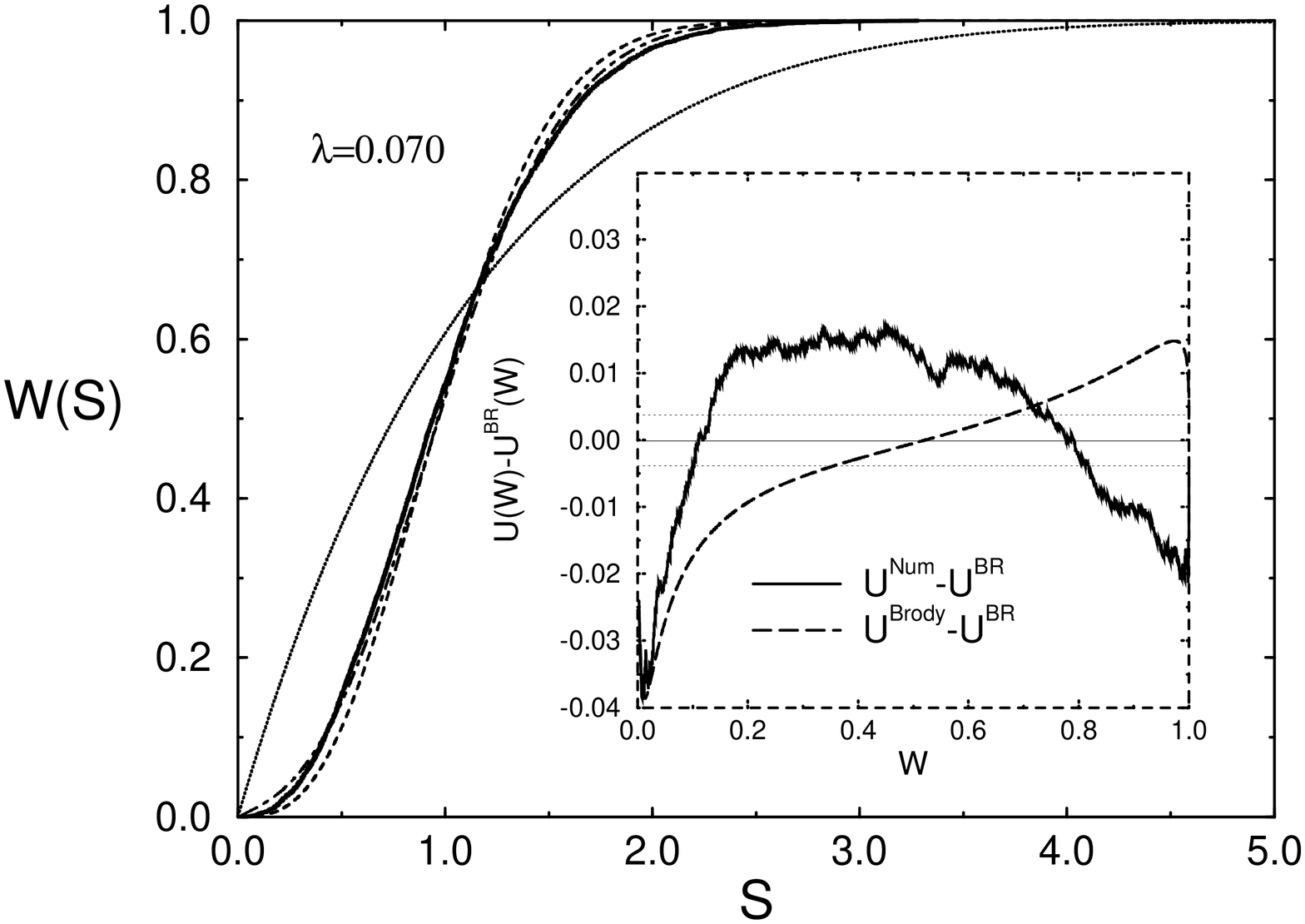,height=8cm,width=3.40cm,angle=-90,
                        bbllx=80,bblly=80,bburx=520,bbury=770}
\end{figure}
\newpage
\begin{figure}[H]
\epsfig{file=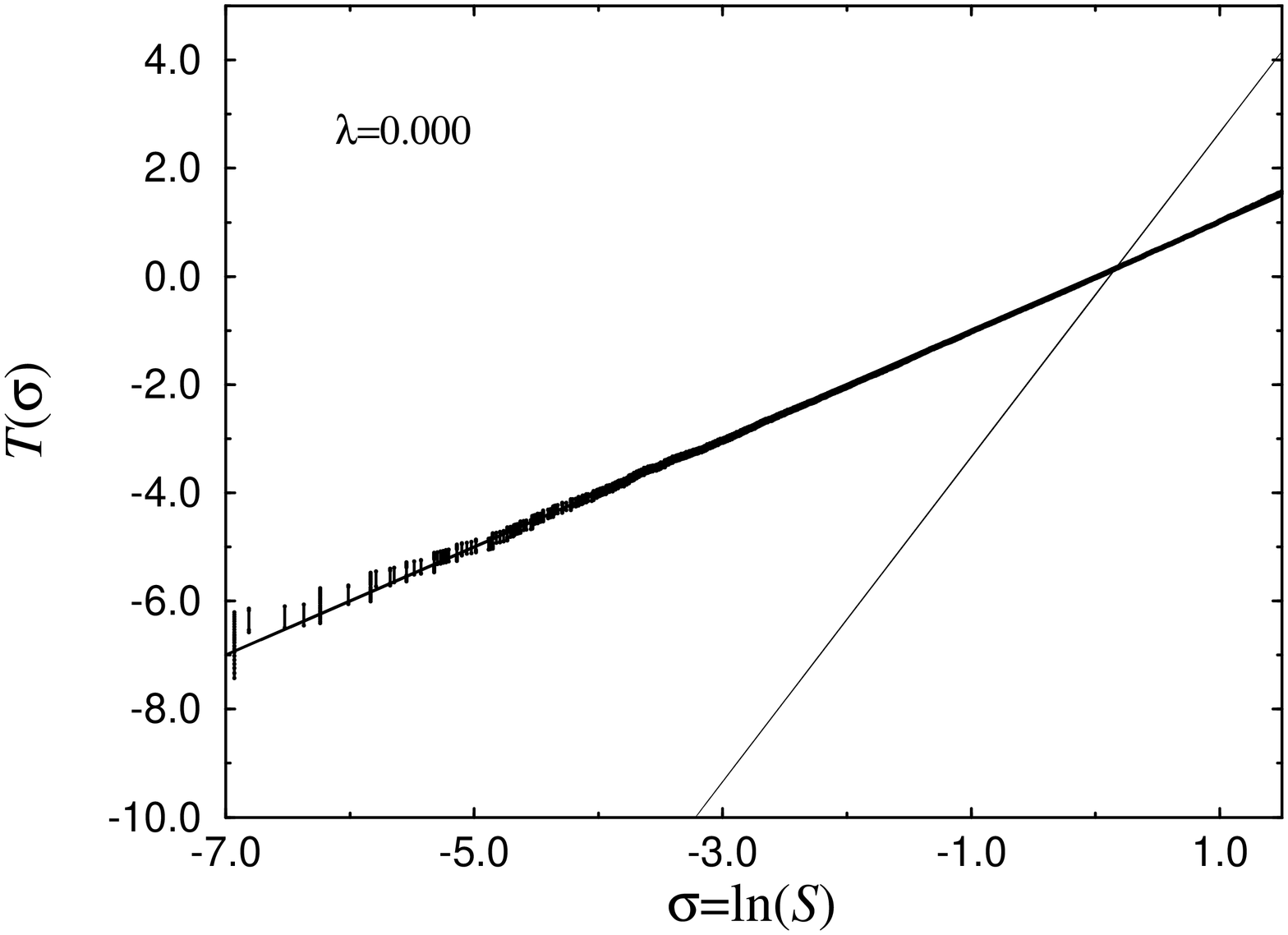,height=8cm,width=3.40cm,angle=-90,
                        bbllx=80,bblly=80,bburx=520,bbury=770}
\end{figure}
\begin{figure}[H]
\epsfig{file=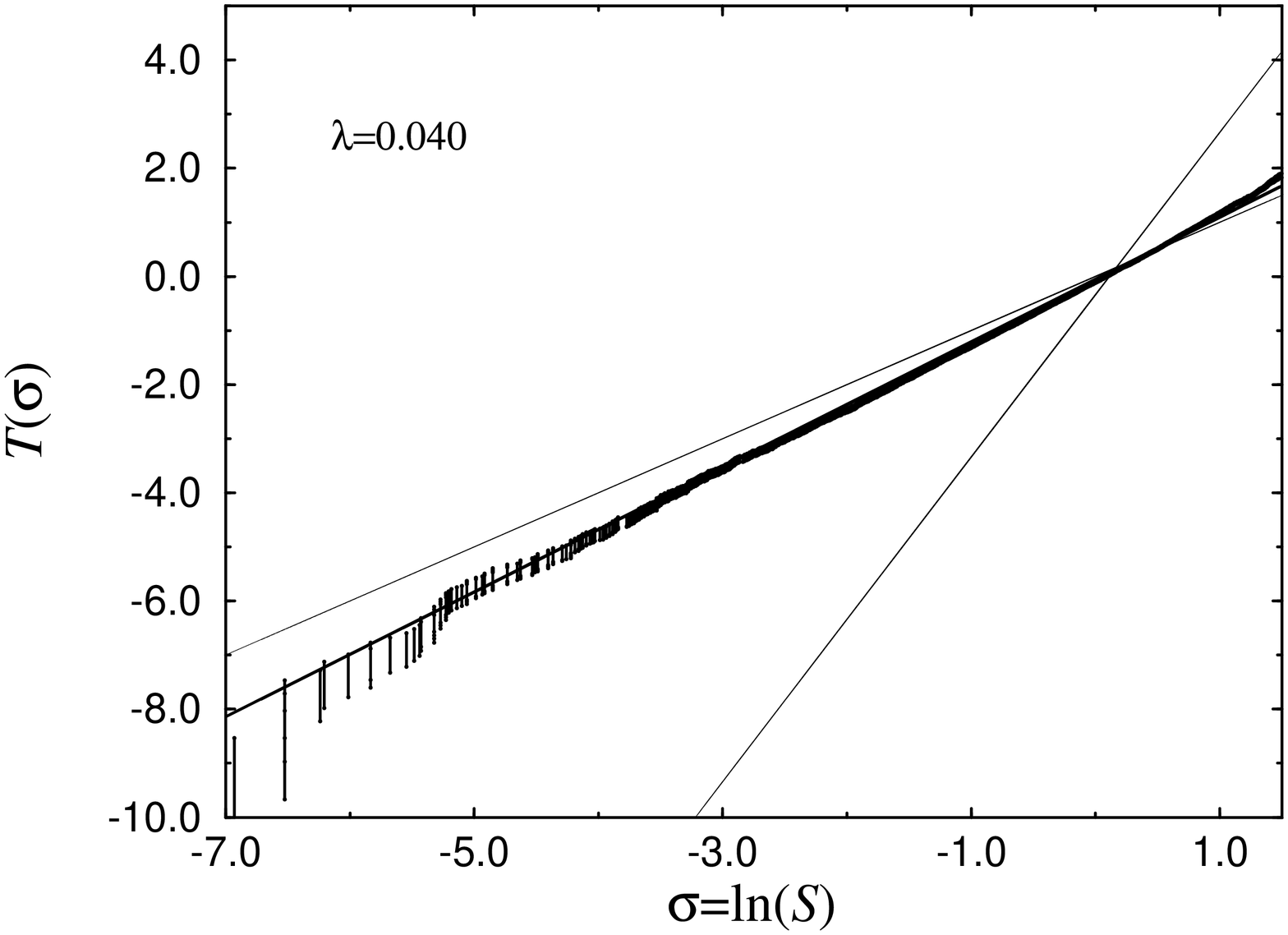,height=8cm,width=3.40cm,angle=-90,
                        bbllx=80,bblly=80,bburx=520,bbury=770}
\end{figure}
\begin{figure}[H]
\epsfig{file=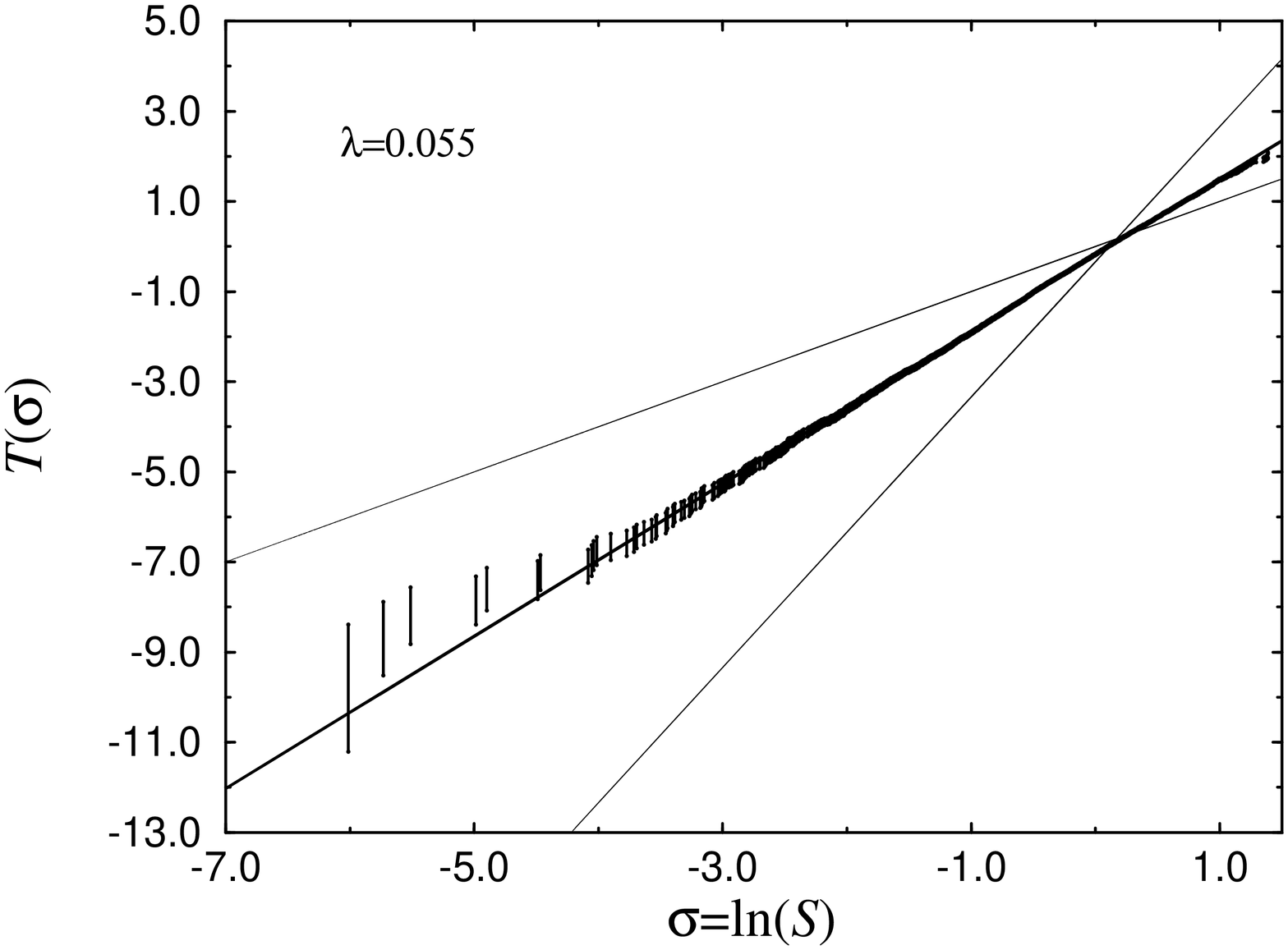,height=8cm,width=3.40cm,angle=-90,
                        bbllx=80,bblly=80,bburx=520,bbury=770}
\end{figure}
\begin{figure}[H]
\epsfig{file=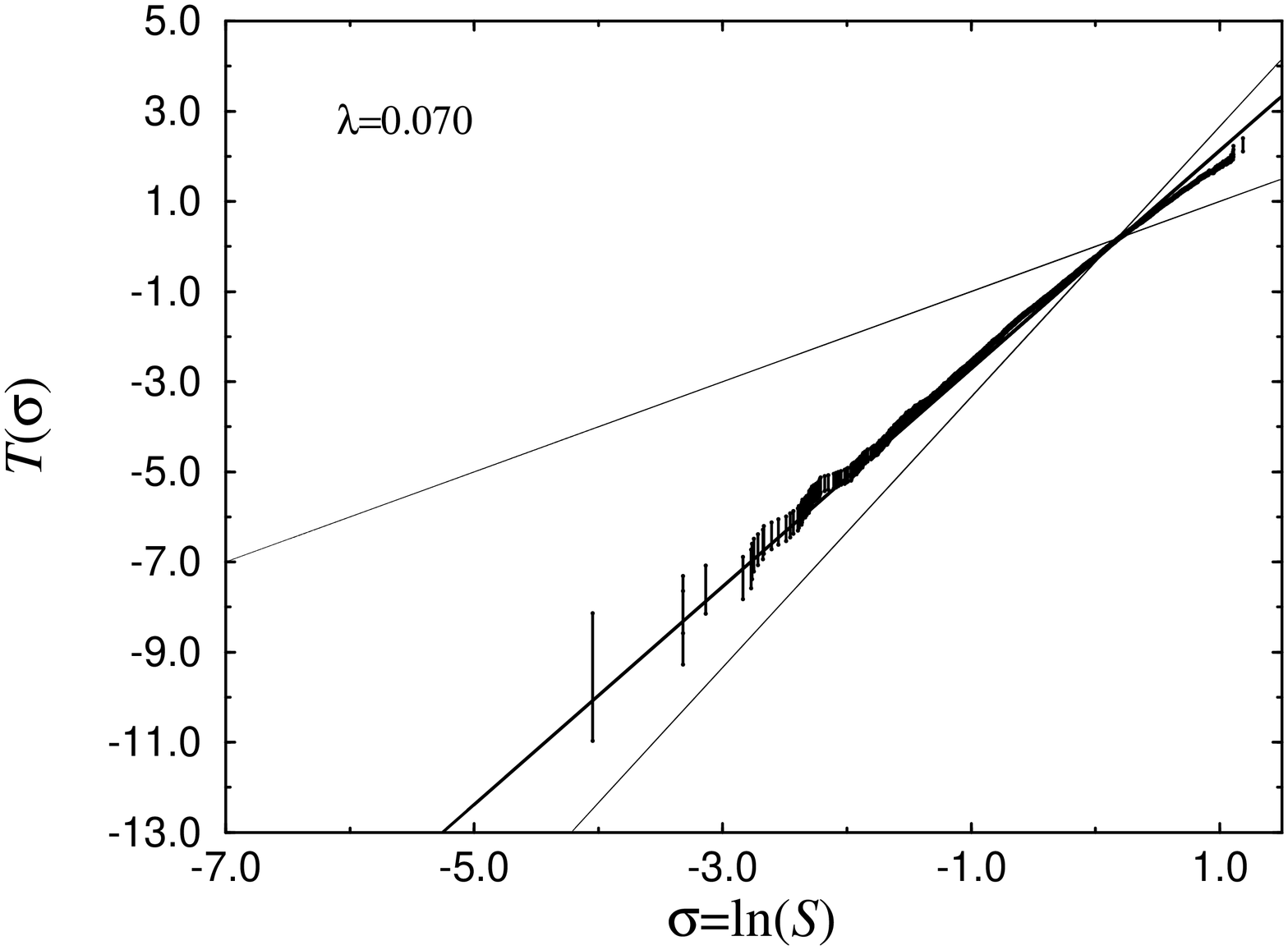,height=8cm,width=3.40cm,angle=-90,
                        bbllx=80,bblly=80,bburx=520,bbury=770}
\end{figure}
\begin{figure}[H]
\epsfig{file=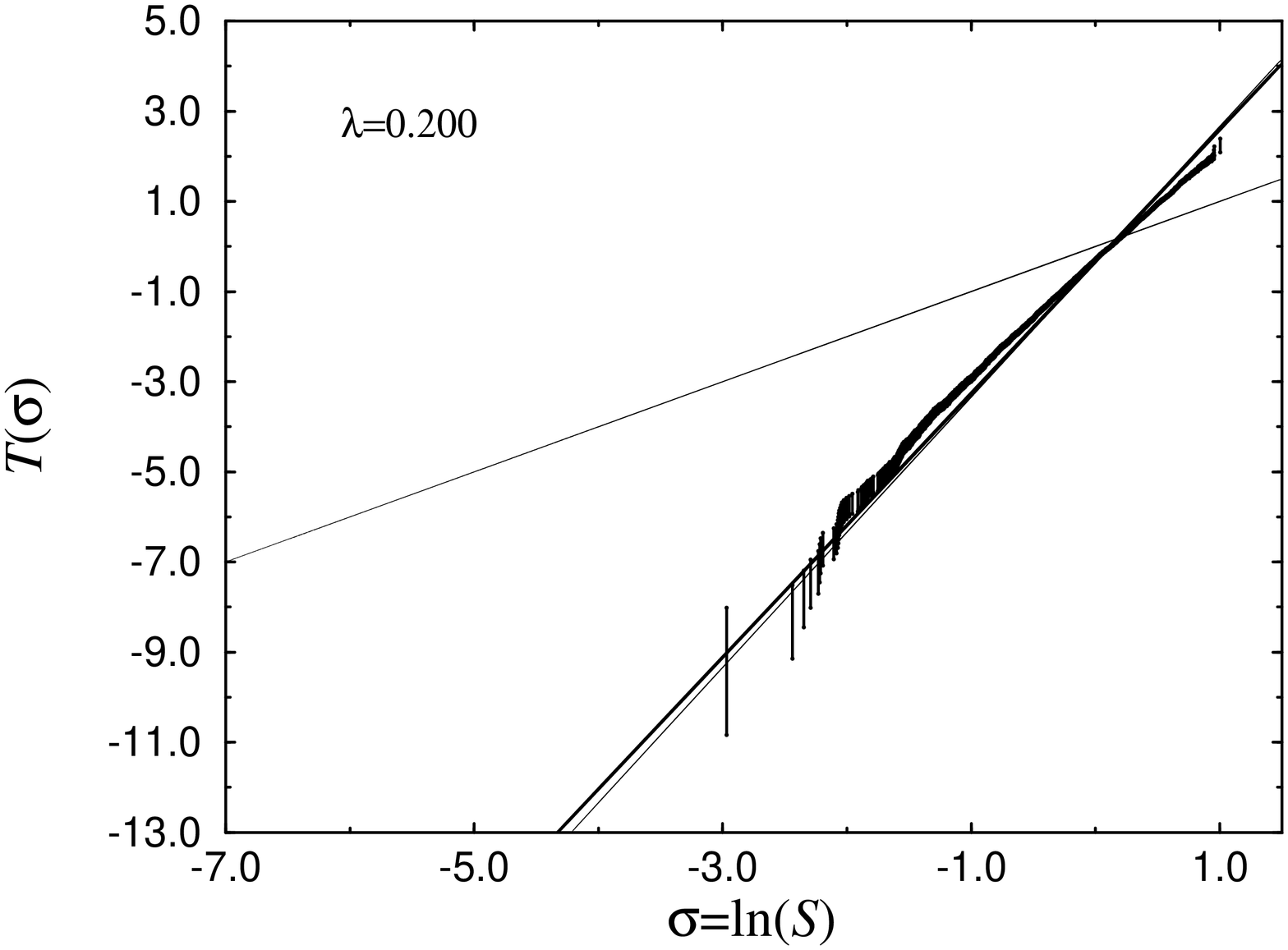,height=8cm,width=3.40cm,angle=-90,
                        bbllx=80,bblly=80,bburx=520,bbury=770}
\vspace{0.5 cm}
\caption{The function $T(\ln{S})$ for some of the calculated spectra. The
  value of the shape parameter is written in the upper left corner of
  each figure.}
\label{TlnS}
\end{figure}
The results are shown in Fig. \ref{TlnS} and in Table \ref{tab0} and
Fig. \ref{final}. In the Fig. \ref{TlnS} there is the $T$ function
shown, which transforms Brody distribution to the straight line:
$T(\ln{S})=\ln\left[-\ln\left\{1-W(\exp(\ln{S}))\right\}\right]$. Fitting
the numerical data to the line, the slope is the exponent of the level
repulsion plus one. The statistical error of the measured values is
shown with the bars.

A good agreement with the Poisson statistics is seen in the integrable
case ($\lambda=0$), while small deviations from GUE statistics in the
fully chaotic case are observed. The deviations at $\lam=0.2$ are
probably due to localization of the wave functions, e.g. scars, and
further work is necessary to explain that. For small $S$ the quadratic
level repulsion is verified at $\lambda=0.2$. The fractional power law
level repulsion with the exponent $\beta$ varying continuously from 0
to 2 is observed. From the Fig. \ref{TlnS} we can also confirm that
the Brody distribution is valid only locally, for $S<1$.

As for the number variance $\Sigma^2(L)$, we have observed continuous
transition from Poisson, $\Sigma^2_{Poisson}(L)=L$, to GUE,
$\Sigma^2_{GUE}(L)\approx 0.10 \ln{L}+0.34$, statistics. The saturation
sets in at $L^*\sim 20$, which is roughly in agreement with Berry's
semiclassical theory \cite{Berry85}. In the fully chaotic case we have
found very nice agreement with the GUE.

We also explored the relevance of the Berry-Robnik theory, which is
based on the statistically independent superposition of Poissonian
sequence of regular levels and GUE sequence of chaotic levels, each
having statistical weight equal to the relative measure of the
corresponding classical invariant component.

Following the procedure in \cite{BR} we derived the Berry-Robnik
formula for Poisson-GUE transition:
\begin{eqnarray}
P^{\rm BR}(S)&=&e^{-\rho_1 S}\,\Bigl\lbrack
\left( {\textstyle {32\over \pi^2}} \rho_2^4 S^2+
{\textstyle {8\over \pi}}\rho_1 \rho_2^2 S+\rho_1^2 \right) 
\,e^{-{4\over \pi}\rho_2^2 S^2}+ \nonumber
\\
&+&\left (2\rho_1\rho_2-\rho_1^2\rho_2 S \right )\,
{\rm erfc}({\textstyle {2\over \sqrt{\pi}}}\rho_2 S) \Bigr\rbrack \;,
\label{BerryRobnik}
\end{eqnarray}
where $\rho_1=1-\rho_2$ is the relative measure of the regular part of
the classical phase space.  The classical work is to determine
$\rho_1^{cl}$ and $\rho_2^{cl}$. Different methods and difficulties
connected to this work are presented in \cite{paper1}. After the
careful study the following method has been chosen.  The SOS
{\footnote{Surface of Section, chosen in the standard way, see
\cite{cubic} and \cite{diploma}}} is first divided into a mesh of
$M\times M$ cells.  A trajectory is then run on the largest chaotic
component. Each cell $i$ has a counter which counts the number of
times a trajectory passes the cell, $\eta_{i}$. After having made
enough steps the distribution of the numbers $\eta$ is plotted. The
distribution has a peak around $\eta=0$, a minimum and another maximum
for larger values of $\eta$.  If the distribution is normalised, the
area to the right of the minimum is equal to $\rho_2$. In table 1
there are values of $\rho_2$ for different shape parameters $\lambda$.
\begin{figure}
\epsfig{file=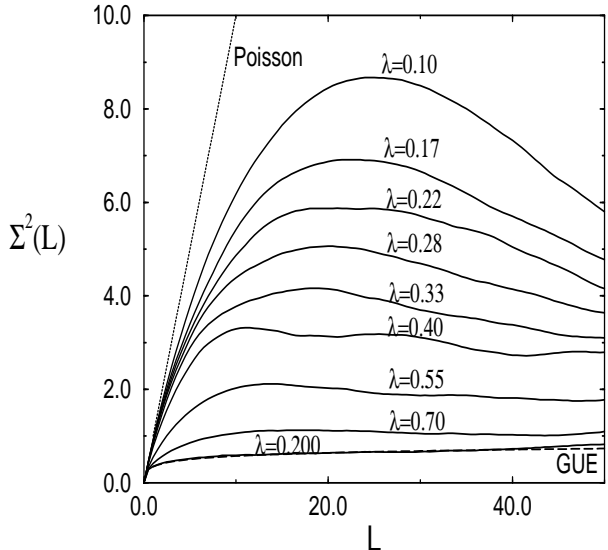,height=8cm,width=7cm,angle=-90,bbllx=80,
				       bblly=80,bburx=520,bbury=770}
\vspace{0.5 cm}
\caption{The number variance statistics $\Sigma^2 (L)$ for different shape 
parameters. The limit cases are shown dotted (Poisson) and dashed (GUE)}
\label{NV}
\end{figure}
\begin{table}
\begin{tabular}{|c||c|c|c|}
$\lambda$&$\rho^{cl}_2$&$\rho^{BR}_2$&$\beta$ \\ \hline\hline
0.000 & 0.00 & 0.18 & 0.00 \\ \hline
0.010 & 0.05 & 0.26 & 0.02 \\ \hline
0.017 & 0.15 & 0.28 & 0.03 \\ \hline
0.022 & 0.26 & 0.32 & 0.05 \\ \hline
0.025 & 0.34 & 0.35 & 0.09 \\ \hline
0.028 & 0.58 & 0.36 & 0.03 \\ \hline
0.033 & 0.75 & 0.41 & 0.15 \\ \hline
0.040 & 0.86 & 0.49 & 0.16 \\ \hline
0.049 & 0.92 & 0.68 & 0.48 \\ \hline
0.055 & 0.99 & 0.78 & 0.69 \\ \hline
0.062 & 0.99 & 0.77 & 0.87 \\ \hline
0.070 & 0.99 & 0.94 & 1.42 \\ \hline
0.100 & 1.00 & 0.97 & 1.58 \\ \hline
0.200 & 1.00 & 0.99 & 1.93
\end{tabular}
\caption{The classical value $\rho_2^{cl}$ for different shape 
parameters $\lambda$. The third and the fourth column are the
best-fitting values for the Berry-Robnik and Brody model, respectively
(see further text).}
\label{tab0}
\end{table}
The quantum parameter $\rho_2^{BR}$ is determined through fitting to
the equation (\ref{BerryRobnik}) and should be compared to the
classical measure $\rho_2^{cl}$. The results are presented in
Fig. \ref{final} and in Table \ref{tab0}. In insets of Fig. \ref{WS}
the fine scale deviations from the Berry-Robnik formula are plotted in
terms of the $U(W(S))-U(W^{BR}(S))$ vs. $W(S)$ (full line). We also
plot $U(W^{Brody}(S))-U(W^{BR}(S))$ for comparison (dashed line). The
transformation $U(W)={\textstyle{2\over\pi}}\arccos{\sqrt{1-W}}$ is
used \cite{PR93} in order to have uniform statistical error over the
plot, $\delta U=1/(\pi\sqrt{N})$. The dotted horizontal lines indicate
$\pm \delta U$. The Berry-Robnik fit is not statistically significant
and the values of the fitting parameters $\rho_2^{\rm BR}$ are
different from the classical values $\rho_2^{\rm cl}$. Why this is so
and how the crossover from Brody-like behaviour to Berry-Robnik takes
place is discussed in \cite{RP97}. As one can see at small $S$ (and
therefore at small $W(S)$) the agreement with Brody is better than
Berry-Robnik, and occasionally, like e.g. at $\lambda=0.055$ and
$0.040$, even globally Brody seems to capture the behaviour of our
data, although, on the other hand, there are {\sl expected deviations}
at large $S$ (and large $W(S)$) e.g. in the fully chaotic case of
Africa billiard ($\lambda=0.2$), where $\beta=1.93$, and Brody
asymptotic behaviour at large $S$ is certainly wrong, because its
exponent varies $\propto S^3$, which is different from GUE in equation
(\ref{GUE}), $\propto S^2$. What we see in this $U$-function
representation is consistent with the $T$-function representation in
Fig. \ref{TlnS}.
\\\\
In this paper we have presented the numerical computation and
statistical analysis of energy spectra of a Hamiltonian system with
mixed classical dynamics and without antiunitary symmetry, namely the
Aharonov-Bohm cubic conformal plane billiard. We reconfirmed the GUE
statistics in the ergodic (fully chaotic) case with approximately 6000
good consecutive energy levels whose accuracy is better than one
percent of the mean level spacing. Further, in the KAM regime we have
found {\sl the fractional power law level repulsion} with the exponent
varying smoothly from zero to two as the corresponding classical
dynamics goes from integrable to ergodic. On the other hand we know
that in the ultimate semiclassical limit the Berry-Robnik theory
\cite{BR,RP97} should apply, which does not exhibit level repulsion
($P^{BR}(S=0)=\rho_1^2\ne 0$). The resolution of this disagreement
lies in the fact that the key assumptions of the Berry-Robnik theory
are fulfilled only for extremely large sequential quantum numbers or
small effective Planck's constant. Namely, one should require
extendedness of the Wigner functions on the whole corresponding
classically allowed invariant components of phase space. This is true
only if the {\sl Heisenberg time}, $t_{H}=2\pi\hbar / \langle\Delta
E\rangle$ (the time until the discreteness of the spectrum is not yet
resolved and the quantum evolution follows the classical one), is
larger than the classical {\sl ergodic time}, $t_e$ (the typical time
after which classical distributions reach equilibrium steady state on
the invariant set). In our case we have typically $t_{e}\gtrsim 10^4$
(velocity is one), so the condition $t_{H}>t_{e}$ yields that the
sequential number $N$ should be larger than about $10^8$. We believe
that our present work contributes to the new developements in the wide
field of research of quantum chaos, recently reviewed by
Weidenm\"uller et al. \cite{GMW}.

\acknowledgements
J.D. would like to thank Darko Veberi\v c for reading the
manuscript. The work was supported by the Ministry of Science and
Technology of the Republic of Slovenia and by the Rector's Funnd of
the University of Maribor.


\begin{thebibliography}{}
\bibitem{antiunit} M.~Robnik Lecture Notes in Physics {\bf 226}, 120 (1986).
\bibitem{cubic} M.~V.~Berry and M.~Robnik J.~Phys.~A:~Math.~Gen.  
{\bf 19}, 649 (1986).
\bibitem{R92} M.~Robnik J.~Phys.~A:~Math.~Gen. {\bf 26} 1399, 3593 (1992).
\bibitem{BGS} O.~Bohigas, M.~J.~Giannoni, and C.~Schmit Phys.~Rev.~Lett. 
{\bf 226}, 120 (1984).
\bibitem{PR94} T.~Prosen and M.~Robnik J.~Phys.~A:~Math.~Gen. 
{\bf 27} 8059 (1994).
\bibitem{PR93} T.~Prosen and M.~Robnik J.~Phys.~A:~Math~.~Gen.  
{\bf 26}, 1105 (1993).
\bibitem{BR} M.~V.~Berry and M.~Robnik J.~Phys.~A:~Math.~Gen.  
{\bf 17} 2413 (1984).
\bibitem{Haake} F.~Haake {\sl Quantum Signatures of Chaos}, 
Berlin: Springer, (1992).
\bibitem{BohTomUl} O.~Bohigas, S.~Tomsovic, and D.~Ullmo 
Physics~Reports, {\bf 223}, No. 2 , 43-133 (1993).  
\bibitem{diploma} J.~Dobnikar {\sl Spectral statistics in the 
transition region between integrability and chaos for systems with
broken antiunitary symmetry}, Diploma thesis, CAMTP and Physics
department, FMF, University of Ljubljana, (1996), unpublished.
\bibitem{R97} M.~Robnik Open~Systems~and~Information~Dynamics
{\bf 4} 211 (1997).
\bibitem{R84} M.~Robnik J.~Phys.~A:~Math.~Gen. {\bf 17} 1049 (1984).
\bibitem{Brody} T.~A.~Brody Lett.~Nuovo~Cimento {\bf 7} 482 (1973).
\bibitem{Berry85} M.~V.~Berry Proc.~Roy.~Soc. London {\bf A400} 229 (1985).
\bibitem{paper1} M.~Robnik, J.~Dobnikar, A.~Rapisarda, T.~Prosen, 
and M.~Petkov\v sek J.~Phys.~A:~Math.~Gen., in press.
\bibitem{RP97} M.~Robnik and T.~Prosen J.~Phys.~A:~Math.~Gen.
{\bf 30} in press (1997).
\bibitem{GMW} T.~Guhr, A.~M\"uller-Groeling and H.~A.~Weidenm\"uller
{\sl Random Matrix Theories in Quantum Physics: Common Concepts},
MPI Preprint {\bf H V27}, MPIfK Heidelberg, cond-mat/9707301, July 1997. 
\end{thebibliography}
\end{document}